\documentclass[dvips]{article}
\usepackage{icrctc07}

\title{MAGIC upper limits on the high energy
emission from GRBs}
\shorttitle{MAGIC Uper Limits on GRBs}
\authors{D. Bastieri$^{1}$, N. Galante$^{2}$, M. Garczarczyk$^{2}$, M. Gaug$^{3}$, 
F. Longo$^{4}$, S. Mizobuchi$^{2}$ and \mbox{V. Scapin}$^{5}$, for the MAGIC collaboration}
\shortauthors{N. Galante et al}
\afiliations{
$^1$ Dipartimento di Fisica - Universit\`a di Padova \& INFN Padova, Via F. Marzolo 8, I-35131 Padova, Italy\\ 
$^2$ Max-Planck-Institut f\"ur Physik, F\"ohringer Ring 6, D-80805 Munich, Germany\\
$^3$ Instituto de Astrof\'isica de Canarias, Calle V\'ia L\'actea, E-38205 La Laguna, Tenerife, Spain\\
$^4$ Dipartimento di Fisica - Universit\`a di Trieste \& INFN Trieste, Via F. Valerio 2, I-34127 Trieste, Italy\\ $^5$ Dipartimento di Fisica - Universit\`a di Udine \& INFN Trieste, Via delle Scienze 208, I-33100 Udine, Italy}

\email{galante@mppmu.mpg.de}

\abstract{During its cycle-1 observation period, between April~2005 and March~2006, the MAGIC telescope was able to observe nine Gamma Ray Burst (GRB) events since their early beginning. Other observations were performed during the following months in the cycle-2 observation period. The observations, with an energy threshold spanning from 80 to $200 \, \mathrm{GeV}$, did not reveal any $\gamma$-ray emission. The computed upper limits are compatible with a power law extrapolation, where intrinsic fluxes are evaluated taking into account the attenuation due to the scattering in the metagalactic radiation field.}

\begin{document}
\maketitle

\section{Introduction}

The Major Atmospheric Gamma Imaging Cherenkov (MAGIC) telescope~\cite{ref1}, located on the Canary Island of La Palma, is currently the largest Imaging Air Cherenkov Telescope (IACT). MAGIC has a $17 \, \mathrm{m}$ diameter tessellated reflector dish consisting of 964 ($0.5 \times 0.5 \mathrm{m}^2$) diamond-milled aluminium mirrors. In its current configuration, the MAGIC photo-multiplier camera has a trigger region of $2.0^\circ$ diameter~\cite{ref2}, and a trigger collection area for $\gamma$-rays of the order of $10^5 \, \mathrm{m}^2$, which increases further with the zenith angle of observation. Presently, the accessible trigger energy ranges from $50-60 \, \mathrm{GeV}$ (at small zenith angles) up to the TeV region close to the horizon. The MAGIC telescope is focused to $10 \, \mathrm{km}$ distance -- the most likely height at which a $50 \, \mathrm{GeV}$ $\gamma$-ray shower has its shower maximum. The accuracy in reconstructing the direction of incoming $\gamma$-rays, the point spread function (PSF), is about $0.1^\circ$, slightly depending on the analysis.

In this contribution we report the results of observation of GRBs by the MAGIC telescope during cycle-1. Preliminary results from a selected sample of GRBs of cycle-2 are also included. The technical details of the MAGIC observations are reported in a separate contribution~\cite{ref3}.

\section{The Observations}

During the fourteen months of observation \mbox{cycle-1} nine GRB events suitable for analysis were selected. Approximately the same amount of events was also collected during cycle-2, spanning from May~2006 up to May~2007, out of which four were analyzed in a preliminary way. Table~\ref{tab:grbs} summarizes the thirteen GRBs observed by MAGIC up to now with their principal features obtained from the GCN circulars. 

\begin{table*}
\begin{center}
\begin{tabular}{c|c|c|r|r|r|c|c}
   & GRB  & Satellite & Trigger \# & Energy Range & $T_{90} $ & Fluence & $z$ \\
\hline
1. & 050421   & {\it Swift} & 115135 & 15-350 keV & 10 s & $1.8\times 10^{-7}$ & -       \\
2. & 050502a   & Integral & 2484 & 20-200 keV & 20 s & $1.4\times 10^{-6}$ & 3.79  \\
3. & 050505     & {\it Swift} & 117504 & 15-350 keV & 60 s & $4.1\times 10^{-6}$ & 4.27 \\
4. & 050509a  & {\it Swift} & 118707 & 15-350 keV & 13 s & $4.6\times 10^{-7}$ & - \\
5. & 050713a & {\it Swift} & 145675 & 15-350 keV & 70 s & $9.1\times 10^{-6}$ & - \\
6. & 050904    & {\it Swift} & 153514 & 15-350 keV & 225 s & $5.4\times 10^{-6}$ & 6.29 \\
7. & 060121   & HETE-2 &  4010 & 0.02-1 MeV & 2 s & $4.7\times 10^{-6}$ & - \\
8. & 060203  & {\it Swift} & 180151 & 15-350 keV & 60 s & $8.5\times 10^{-7}$ & - \\
9. & 060206  & {\it Swift} & 180455 & 15-350 keV & 11 s & $8.4\times 10^{-7}$ & 4.05 \\
10. & 060825 & {\it Swift} & 226382 & 15-350 keV & 15 s & $9.8\times 10^{-7}$ & - \\
11. & 061028 & {\it Swift} & 235715 & 15-350 keV & 106 s & $9.7\times 10^{-7}$ & - \\
12. & 061217 & {\it Swift} & 251634 & 15-350 keV & 0.4 s & $4.6\times 10^{-8}$ & 0.83 \\
13. & 070412 & {\it Swift} & 275119 & 15-350 keV & 40 s & $4.8\times 10^{-7}$ & - \\
\end{tabular}\\
\end{center}
\caption{\label{tab:grbs} Main properties of GRBs observed by MAGIC. The fourth column shows the typical energy range of the detector on board of the GRB triggering satellite, while the fifth, sixth and seventh columns give the corresponding measured duration $T_{90}$, fluence in [erg~cm$^{-2}$] and redshift.}
\end{table*}

The majority of the events were triggered by the {\it Swift} satellite. GRB050502a and GRB060121 were triggered by Integral and \mbox{HETE-2} respectively. Apart from GRB060121 and GRB061217, all GRBs observed by MAGIC are long bursts. In two cases, for GRB050713a and GRB050904, the fast repositioning of the telescope allowed the observation of their prompt emission phase~\cite{ref10,ref3}, while the burst was still emitting low-energy $\gamma$-rays. 

In six cases the early GRB afterglow observation with MAGIC overlapped with the X-ray observation from space, namely for GRB050421, GRB050713a, GRB050904, GRB060206, GRB060825 and GRB061028. During cycle-1, in the case of GRB050421 the X-Ray Telescope (XRT) on board {\it Swift} detected two flaring events in the afterglow at $T_0 + 110 \, \mathrm{s}$ and $T_0 + 154 \, \mathrm{s}$, as well as in GRB050904, where a clear flare was detected at $T_0 + 466 \, \mathrm{s}$. In these two cases data taking with the MAGIC telescope started at $T_0 + 108 \, \mathrm{s}$ and $T_0 + 145 \, \mathrm{s}$ respectively, the flaring activity in X-rays was therefore covered by the MAGIC observation window. During cycle-2 only in the case of GRB060825 flaring X-ray emission was also covered by the MAGIC observation window. 

\section{Analysis and Results}

For the reconstruction of $\gamma$-events the standard MAGIC analysis software~\cite{ref4} based on the Hillas analysis~\cite{ref5} was used. The reconstructed signal was calibrated and cleaned from spurious backgrounds using an image cleaning algorithm which requires a signal exceeding fixed reference levels, improved with the reconstructed information of the arrival time~\cite{ref6}. Some additional quality cuts based on combinations of Hillas parameters were applied to remove non-physical images from the data. The separation between $\gamma$ and hadron showers was performed by means of the Random Forest (RF) method, a classification method that combines several parameters describing the shape of the image into a new parameter called \mbox{\emph{hadronness}}~\cite{ref7}. The energy of the $\gamma$-ray was also estimated using the RF approach, yielding a resolution of $\sim30\%$ at $200 \, \mathrm{GeV}$.

Dedicated OFF-data samples were selected for each GRB according to the same observation conditions (zenith angle, discriminator threshold, trigger rate, Moon phase). The cut on \emph{hadronness}, ensuring at least 90\% of efficiency on $\gamma$-like events, was optimized using a dedicated Monte Carlo simulation. In the final analysis step a cut on the \emph{alpha} parameter was chosen. This parameter is related to the direction of the incoming shower, thus it is expected to peak at $alpha =0^\circ$ if the telescope points directly at the source, while it is uniformly distributed for background events. The \emph{alpha} parameter is used to evaluate the significance of the signal.

\begin{table*}[!hbp]
\begin{center}
\begin{tabular}{l|c|c|c@{\quad}c@{\quad}c}
{} & \textbf{Energy bin} & \textbf{Energy} & \multicolumn{3}{c}{\textbf{Fluence Upper Limit}} \\
{} & \textbf{[GeV]} & \textbf{[GeV]} & \textbf{[cm$^{-2}$ keV$^{-1}$]} & \textbf{[erg cm$^{-2}$]} & \textbf{C.U.} \\
\hline
\hline
{} & 175-225 & 212.5 & $5.26\times 10^{-16}$ & $3.80\times 10^{-8}$ & 0.20 \\
\cline{2-6}
{} & 225-300 & 275.8 & $3.64\times 10^{-16}$ & $4.43\times 10^{-8}$ & 0.27 \\
\cline{2-6}
\raisebox{1.5ex}[-1.5ex]{\textbf{GRB050421}} & 300-400 & 366.4 & $5.21\times 10^{-17}$ & $1.12\times 10^{-8}$ & 0.08 \\
\cline{2-6}
{} & 400-1000 & 658.7 & $2.07\times 10^{-17}$ & $1.41\times 10^{-8}$ & 0.14 \\
\hline
\hline
{} & 120-175 & 152.3 & $1.67\times 10^{-15}$ & $6.21\times 10^{-8}$ & 0.27 \\
\cline{2-6}
{} & 175-225 & 219.3 & $2.83\times 10^{-15}$ & $2.18\times 10^{-7}$ & 1.15 \\
\cline{2-6}
\textbf{GRB050502a} & 225-300 & 275.8 & $1.13\times 10^{-15}$ & $1.37\times 10^{-7}$ & 0.83 \\
\cline{2-6}
{} & 300-400 & 360.8 & $7.57\times 10^{-17}$ & $1.58\times 10^{-8}$ & 0.11 \\
\cline{2-6}
{} & 400-1000 & 629.1 & $5.62\times 10^{-17}$ & $3.56\times 10^{-8}$ & 0.35 \\
\hline
\hline
{} & 175-225 & 212.9  & $2.03\times 10^{-15}$  & $1.48\times 10^{-7}$ & 0.76  \\
\cline{2-6}
{} & 225-300 & 275.1 & $2.66\times 10^{-15}$ & $3.22\times 10^{-7}$ & 1.94 \\
\cline{2-6}
\raisebox{1.5ex}[-1.5ex]{\textbf{GRB050505}} & 300-400 & 363.6 & $5.28\times 10^{-16}$ & $1.11\times 10^{-7}$ & 0.79 \\
\cline{2-6}
{} & 400-1000 & 704.1 & $1.85\times 10^{-17}$ & $1.46\times 10^{-8}$ & 0.15 \\
\hline
\hline
{} & 175-225 & 215.1 & $1.04\times 10^{-15}$ & $7.69\times 10^{-8}$ & 0.40 \\
\cline{2-6}
{} & 225-300 & 273.4 & $1.39\times 10^{-15}$ & $1.67\times 10^{-7}$ & 1.00 \\
\cline{2-6}
\raisebox{1.5ex}[-1.5ex]{\textbf{GRB050509a}} & 300-400 & 362.8 & $7.74\times 10^{-16}$ & $1.63\times 10^{-7}$ & 1.15\\
\cline{2-6}
{} & 400-1000 & 668.5 & $1.69\times 10^{-16}$ & $1.21\times 10^{-7}$ & 1.22\\
\hline
\hline
{} & 120-175 & 169.9 & $3.63\times 10^{-15}$ & $1.68\times 10^{-7}$ & 0.76 \\
\cline{2-6}
{} & 175-225 & 212.5 & $1.12\times 10^{-15}$ & $8.08\times 10^{-8}$ & 0.42 \\
\cline{2-6}
\textbf{GRB050713a} & 225-300 & 275.8 & $2.07\times 10^{-15}$ & $2.52\times 10^{-7}$ & 1.52 \\
\cline{2-6}
{} & 300-400 & 366.4 & $3.33\times 10^{-16}$ & $7.16\times 10^{-8}$ & 0.51 \\
\cline{2-6}
{} & 400-1000 & 658.7 & $2.24\times 10^{-17}$ & $1.55\times 10^{-8}$ & 0.15 \\
\hline
\hline
{} & 80-120 & 85.5 & $9.06\times 10^{-15}$ & $1.06\times 10^{-7}$ & 0.32 \\
\cline{2-6}
{} & 120-175 & 140.1 & $3.00\times 10^{-15}$ & $9.42\times 10^{-8}$ & 0.38 \\
\cline{2-6}
{} & 175-225 & 209.9 & $2.18\times 10^{-15}$ & $1.53\times 10^{-7}$ & 0.79 \\
\cline{2-6}
\raisebox{1.5ex}[-1.5ex]{\textbf{GRB050904}} & 225-300 & 268.9 & $5.82\times 10^{-16}$ & $6.74\times 10^{-8}$ & 0.40 \\
\cline{2-6}
{} & 300-400 & 355.2 & $5.01\times 10^{-16}$ & $1.11\times 10^{-7}$ & 0.71 \\
\cline{2-6}
{} & 400-1000 & 614.9 & $1.26\times 10^{-16}$ & $7.63\times 10^{-8}$ & 0.73 \\
\hline
\hline
{} & 120-175 & 151.3 & $2.64\times 10^{-15}$ & $9.67\times 10^{-8}$ & 0.41 \\
\cline{2-6}
{} & 175-225 & 212.8 & $6.57\times 10^{-16}$ & $4.76\times 10^{-8}$ & 0.25 \\
\cline{2-6}
\textbf{GRB060121} & 225-300 & 273.7 & $2.13\times 10^{-16}$ & $2.56\times 10^{-8}$ & 0.15 \\
\cline{2-6}
{} & 300-400 & 367.7 & $4.47\times 10^{-16}$ & $9.66\times 10^{-8}$ & 0.69 \\
\cline{2-6}
{} & 400-1000 & 636.4 & $4.84\times 10^{-17}$ & $3.14\times 10^{-8}$ & 0.31 \\
\hline
\hline
{} & 120-175 & 151.5 & $1.10\times 10^{-14}$ & $4.03\times 10^{-7}$ & 1.71 \\
\cline{2-6}
{} & 175-225 & 219.5 & $5.07\times 10^{-16}$ & $3.91\times 10^{-8}$ & 0.21 \\
\cline{2-6}
\textbf{GRB060203} & 225-300 & 274.0 & $1.57\times 10^{-16}$ & $1.88\times 10^{-8}$ & 0.11 \\
\cline{2-6}
{} & 300-400 & 365.3 & $3.54\times 10^{-16}$ & $7.56\times 10^{-8}$ & 0.54\\
\cline{2-6}
{} & 400-1000 & 639.5 & $4.45\times 10^{-17}$ & $2.91\times 10^{-8}$ & 0.29 \\
\hline
\hline
{} & 80-120 & 85.5 & $1.23\times 10^{-14}$ & $1.44\times 10^{-7}$ & 0.44 \\
\cline{2-6}
{} & 120-175 & 139.9 & $9.83\times 10^{-16}$ & $3.08\times 10^{-8}$ & 0.13 \\
\cline{2-6}
{} & 175-225 & 210.3 & $5.50\times 10^{-16}$ & $3.89\times 10^{-8}$ & 0.20 \\
\cline{2-6}
\raisebox{1.5ex}[-1.5ex]{\textbf{GRB060206}} & 225-300 & 269.2 & $3.65\times 10^{-16}$ & $4.23\times 10^{-8}$ & 0.25 \\
\cline{2-6}
{} & 300-400 & 355.4 & $6.47\times 10^{-16}$ & $1.31\times 10^{-7}$ & 0.91 \\
\cline{2-6}
{} & 400-1000 & 614.0 & $2.88\times 10^{-17}$ & $1.74\times 10^{-8}$ & 0.17 \\
\hline
\hline
\end{tabular}
\caption{Fluence upper limits on the VHE $\gamma$-ray emission from GRBs observed with the MAGIC telescope. The limits were extracted from the first $30 \, \mathrm{min}$ of observation. The first column show the reconstructed energy range. The second column shows the energy at which the upper limit was calculated. The last column shows the upper limit value in Crab Units ($\mathrm{C.U.} = 1.5 \cdot 10^{-3} \left( \frac{E}{\mathrm{GeV}} \right)^{-2.59} \frac{\mathrm{ph}}{\mathrm{cm^{2} \, s \, TeV}}$).}
\label{tab:ul}
\end{center}
\end{table*}

No significant excess of VHE $\gamma$-rays could be extracted with the standard analysis approach. In parallel to the analysis of the complete data set a search for short emission periods in different time bins was performed. Also with this approach no significant deviation of the number of excess events in comparison to the background was found. The upper limits presented in table~\ref{tab:ul} cover the cycle-1 observations~\cite{ref9}. The results were obtained for the first $30 \, \mathrm{min}$ of MAGIC observation using the Rolke approach~\cite{ref8}, including an estimated systematic uncertainty of $30\%$ on the absolute flux level. Table~\ref{tab:ul2} contains the preliminary upper limits on four GRBs observed during cycle-2, as published in the GCN circulars.

\begin{table}
\begin{center}
\begin{tabular}{l|c|c}
{} & \textbf{Energy bin} & \textbf{Flux U.L.} \\
{} & \textbf{[GeV]} & \textbf{[erg cm$^{-2}$ s$^{-1}$]} \\
\hline
\hline
{} & 80-125 & $1.8\times 10^{-10}$ \\
\cline{2-3}
{} & 125-175 & $1.9\times 10^{-10}$ \\
\cline{2-3}
\raisebox{1.5ex}[-1.5ex]{\textbf{GRB060825}} & 175-300 & $1.2\times 10^{-10}$ \\
\cline{2-3}
{} & 300-1000 & $0.6\times 10^{-10}$ \\
\hline
\hline
{} & 100-125 & $1.6\times 10^{-10}$ \\
\cline{2-3}
{} & 125-175 & $0.8\times 10^{-10}$ \\
\cline{2-3}
\raisebox{1.5ex}[-1.5ex]{\textbf{GRB061028}} & 175-300 & $0.7\times 10^{-10}$ \\
\cline{2-3}
{} & 300-1000 & $0.3\times 10^{-10}$ \\
\hline
\hline
{} & 300-500 & $0.53\times 10^{-10}$ \\
\cline{2-3}
\raisebox{1.5ex}[-1.5ex]{\textbf{GRB061217}} & 500-1000 & $0.35\times 10^{-10}$ \\
\hline
\hline
{} & 80-125 & $1.03\times 10^{-11}$ \\
\cline{2-3}
{} & 125-175 & $0.31\times 10^{-11}$ \\
\cline{2-3}
\raisebox{1.5ex}[-1.5ex]{\textbf{GRB070412}} & 175-300 & $0.75\times 10^{-11}$ \\
\cline{2-3}
{} & 300-1000 & $0.77\times 10^{-11}$ \\
\hline
\hline
\end{tabular}
\caption{Preliminary flux upper limits on four GRBs observed by MAGIC during cycle-2. The first column shows the reconstructed energy bin.
For the bursts GRB060825, GRB061028 and GRB061217 the upper limits correspond to the first $30 \, \mathrm{min}$, while in the case of GRB070412 the whole data set of $2 \, \mathrm{h}$ was used.}
\label{tab:ul2}
\end{center}
\end{table}

\section{Conclusions}

We presented the results of thirteen GRB observations performed by MAGIC during the last two years of operation. As a response to the GCN alerts provided by dedicated satellites, MAGIC was able to start the observation during the early afterglow emission phases, while the X-ray emission was still visible. In three cases the MAGIC observation covered flaring activity in X-rays. In two cases the observation started even when the prompt emission in $\gamma$-rays was still active. No significant $\gamma$-ray emission above $\sim 100 \, \mathrm{GeV}$ was detected. We derived upper limits for the $\gamma$-ray flux between 85 and $1000 \, \mathrm{GeV}$. These limits are compatible with a naive extension of the power law spectrum, when the redshift is known, up to hundreds of GeV. For the first time a IACT is able to perform 
direct rapid observations of the prompt emission phase of GRBs. This is particularly of interest to extend the observational energy range in the so called ``{\it Swift} era'' and for the future experiments as AGILE and GLAST.  

\section{Acknowledgements}

The construction of the MAGIC telescope was mainly made possible by the support of the German BMBF and MPG,
the Italian INFN, and the Spanish CICYT, to whom goes our grateful acknowledgement. We would also like to thank the IAC for the excellent working conditions at the Observatorio del Roque de los Muchachos in La Palma. This work was further supported by ETH Research Grant TH~34/04~3 and the Polish MNiI Grant 1P03D01028.


\end{document}